%% file: main.tex
\documentclass[trackchanges,twocolumn]{aastex7}

\usepackage{amsfonts}
\usepackage{amssymb}
\usepackage{amsthm}
\usepackage{amsmath}
\usepackage{alltt}
\usepackage{epstopdf}
\usepackage{graphicx}
\usepackage{float}
\usepackage{verbatim} 
\usepackage{natbib}
\usepackage{bm}
\usepackage{booktabs} 
\usepackage{multirow} 

\input Tables.tex
\input Figures.tex

\begin{document}

\title{The Effects of Radially Varying Diffusivities on Stellar Convection Zone Dynamics}

\author[orcid=0009-0001-2001-0535,sname='North America']{Brandon J. Lazard}
\affiliation{Department of Earth, Planetary, and Space Sciences, University of California, Los Angeles, CA, USA}

\email[show]{bjlazard@g.ucla.edu}

\author[orcid=0000-0002-2256-5884]{Nicholas A. Featherstone} 
\affiliation{Southwest Research Institute, Department of Solar and Heliospheric Physics, Boulder, CO 80302, USA}
\email{nicholas.featherstone@swri.org}

\author[orcid=0000-0002-8642-2962]{Jonathan M. Aurnou}
\affiliation{Department of Earth, Planetary, and Space Sciences, University of California, Los Angeles, CA, USA}
\email{aurnou@ucla.edu}


\begin{abstract}
\noindent Convection is ubiquitous in stellar and planetary interiors where it likely plays an integral role in the generation of magnetic fields. As the interiors of these objects remain hidden from direct observation, numerical models of convection are an important tool in the study of astrophysical dynamos. In such models, unrealistic large values of the viscous ($\nu$) and thermal ($\kappa$) diffusivity are routinely used as an ad-hoc representation of the effects of subgrid scale turbulence which is otherwise too small-scale to resolve numerically. However, the functional forms of these diffusion coefficients can vary greatly between studies, complicating efforts to compare between results and against observations. We explore this issue by considering a series of non-rotating, non-magnetic, solar-like convection models with varying radial functions for the diffusivities and differing boundary conditions. We find that the bulk kinetic energy scales similarly regardless of the diffusivity parameterization.  This scaling is consistent with a free-fall scaling, wherein viscosity plays a subdominant role in the force balance.  We do not, however, observe such diffusion-free behavior in the convective heat transport.  Our results also indicate that the functional form adopted for the diffusion coefficients can impact the distribution of turbulence within the convective shell. These results suggest that some care should be taken when comparing solar convection models directly against helioseismic observations. 
\end{abstract}


\input Introduction.tex

\section{Numerical Method}
The current study is based on a series of three-dimensional (3D), nonlinear stellar convection models made using the pseudo-spectral convection code \texttt{Rayleigh} \citep{Featherstone2016a, Matsui2016, Featherstone2024}.  We employ a spherical geometry and represent the horizontal variation of all variables along spherical surfaces using spherical harmonics $Y_{\ell}^m(\theta, \phi)$, where $\ell$ is the spherical harmonic degree, and $m$ is the azimuthal mode order. In the radial direction we employ a Chebyshev collocation method, expanding all variables in terms of Chebyshev polynomials $T_n(r)$, where $n$ is the Chebyshev polynomial degree \cite[e.g.,][]{trefethen2000}.

Our study is concerned with convection in the deep stellar interior, well below the photosphere. In this region, plasma motions are subsonic, and perturbations to thermodynamic variables are small with respect to their mean, horizontally-averaged values. Under these conditions, the anelastic approximation provides an appropriate description of the system's compressibility and thermodynamics \citep{Gough1969, Gilman1977}. Within this framework, the continuity equation is given by
\begin{equation}
    \bm{\nabla} \cdot (\overline{\rho} \bm{{u}}) = 0,
\end{equation}
where $\rho$ is density and $\bm{u}$ is the fluid velocity.  Here we adopt the notation that an overbar denotes a time- and horizontally-averaged background state variable that is soley a function of radius.  The momentum equation is given by
\begin{equation}
    \frac{D \bm{u}}{Dt} = -\bm{\nabla} \frac{P}{\overline{\rho}} - \frac{S}{c_p} \bm{g} + \frac{1}{\overline{\rho}} \bm{\nabla} \cdot \mathcal{D},
\end{equation}
where P is the pressure, S is the entropy, $c_p$ is the specific heat at constant pressure, and $\bm{g}$ is the gravitational acceleration.  The viscous stress tensor, $\mathcal{D}$, is defined as
\begin{equation}
    \mathcal{D}_{ij} = 2 \overline{\rho} \nu [e_{ij} - \frac{1}{3} \bm{\nabla} \cdot \bm{{u}}\, \delta_{ij}].
\end{equation}
Here, $e_{ij}$ is the stress tensor, $\delta_{ij}$ is the Kronecker delta, and $\nu$ is a kinematic viscosity. The thermal energy equation is given by
\begin{equation}
    \label{eq:heat}
    \overline{\rho} \overline{T} \frac{DS}{Dt} = \bm{\nabla} \cdot (\overline{\rho}\overline{T} \kappa \bm{\nabla} S) + Q(r) + \Phi,
\end{equation}
where $\overline{T}$ is the background temperature, $\kappa$ is the thermal diffusivity, and $Q(r)$ is an internal heat source.  The viscous heating term, $\Phi$, is given by
\begin{equation}
\label{eq:vischeat}
\Phi=2 \overline{\rho} \nu [e_{ij}e_{ij} - \frac{1}{3}(\bm{\nabla} \cdot \bm{{u}})^2].
\end{equation}
Finally, we adopt an ideal gas 
equation of state such that
\begin{equation}
\overline{P} = \mathcal{R} \overline{\rho} \overline{T},
\end{equation}
which can be linearized such that
\begin{equation}
    \frac{\rho}{\overline{\rho}} = \frac{P}{\overline{P}} - \frac{T}{\overline{T}} = \frac{P}{\gamma \overline{P}} - \frac{S}{c_p}
\end{equation}
where $\mathcal{R}=c_p-c_v$ is the gas constant and $\gamma=c_p/c_v$ is the adiabatic index of the gas, with $c_v$ denoting the specific heat at constant volume.

\section{Numerical Framework}
We have constructed a total of 33 model stellar convection zones designed to study how the convective kinetic energy and heat transport depend on the functional form of thermal and viscous diffusion.  Each model is constructed using a polytropic background state following \cite{JONES2011}, with the polytropic parameters chosen to produce a background state resembling the inner four density scale heights of the solar convection zone as described in \cite{Featherstone2016a}.  The polytropic parameters are summarized in Table \ref{table:poly}.

\tablepoly

\subsection{Boundary Conditions and Initialization}
Each simulation in this study is initialized with zero velocity and a small, random thermal perturbation.  It is then evolved for at least one viscous diffusion time once the kinetic energy of the system achieves a statistically-steady state.   All models considered in this study employ stress-free and impenetrable boundaries such that
\begin{equation}
    \frac{\partial}{\partial r} \left( \frac{v_\theta}{r} \right) = \frac{\partial}{\partial r} \left( \frac{v_\phi}{r} \right) = v_r = 0
\end{equation}
at $r=r_i$ and $r=r_o$.  For the specific entropy, we consider three combinations of boundary conditions and internal heating.  The bulk of our models are constructed such that
\begin{equation}
\frac{\partial S}{\partial r}\bigg\rvert_{r=r_i} = 0\,\,,\,\, \frac{\partial S}{\partial r}\bigg\rvert_{r=r_o} = \beta,
\end{equation}
with the constant $\beta$ chosen so that the emergent conductive flux through the upper boundary balances the energy deposited through the heat source $Q(r)$.  Under this constraint, we have
\begin{equation}
\beta = \frac{1}{\overline{\rho(r_o)}\,\overline{T(r_o)}\,\kappa(r_o)\,c_P\,r_o^2}\int_{r_i}^{r_o} Q(r) r^2 \, dr.
\end{equation}
For $Q$, we adopt the form
\begin{equation}
    Q(r, \theta, \phi) = A (\overline{P}(r) - \overline{P}(r_o)).
\end{equation}
The normalization constant $A$ is defined such that
\begin{equation}
   \label{eq:lstar}
    L_{\star} = 4 \pi \int_{r_i}^{r_o} Q(r) r^2 \, dr ,
\end{equation}
where $L_\ast$ is the stellar luminosity.   As discussed in \citet{Featherstone2016a}, this form of $Q(r)$ closely resembles the radiative heating profile of the Sun for the polytropic parameters considered here.  The thermal energy flux $F(r)$ that convection and conduction must transport across a spherical surface at radius r is then given by
\begin{equation}
    \label{eq:radheat}
    F(r) = \frac{1}{r^2} \int_{r_i}^{r} Q(x) x^2 \, dx .
\end{equation}

In addition to our internally-heated models, a subset of our models were conducted with either fixed-entropy or fixed-flux boundary conditions such that
\begin{equation}
S(r_i) = \Delta S\,\,,\,\, S(r_o)=0
\end{equation}
and
\begin{equation}
\frac{\partial S}{\partial r}\bigg\rvert_{r=r_i} = \Gamma\,\,,\,\, \frac{\partial S}{\partial r}\bigg\rvert_{r=r_o} = \Gamma\frac{r_i^2}{r_o^2},
\end{equation}
where the entropy contrast, $\Delta S$, and the entropy gradient, $\Gamma$, were varied across these two series of models.

\subsection{Diffusion Coefficients}
For all models presented in this study, the viscous and thermal diffusivities possess the form of
\begin{equation}
    \nu, \kappa \propto \rho^{\alpha},
\end{equation}
where several values of the exponent $\alpha$ are considered.  For the internally heated models, we considered  $\alpha$ = \{-1, -0.5, 0, 0.5\}.  For fixed-flux and fixed-entropy models with no internal heating, we considered $\alpha=0$ only.  Profiles of viscosity for these different values of $\alpha$ are illustrated in Figure \ref{fig:viscosity}.

\viscfig

\subsection{Nondimensional Control Parameters}
All models considered in this study are characterized by two nondimensional control parameters.   The first is the Prandlt number, Pr, given by
\begin{equation}
\mathrm{Pr} = \frac{\nu}{\kappa},
\end{equation}
which is set to unity for all models in this study.  Additionally, all models are characterized by a Rayleigh number which describes the relative importance of buoyancy and diffusion.
The thermal energy flux across the convection zone is fixed in the majority of the numerical models carried out here, and so we calculate a bulk, flux-based Rayleigh Number, $\mathrm{Ra_F}$,  for those models.  It is given by
\begin{equation}
\label{eq:RaF}
    \mathrm{Ra_F} = \frac{\widetilde{g} \widetilde{F} H^4}{c_p \widetilde{\rho} \widetilde{T} \widetilde{\nu} \widetilde{\kappa}^2},
\end{equation}
where $H = r_o - r_i$ is the shell depth, $F$ is the thermal energy flux, and tildes represent volume-averaged quantities over the full shell. This formulation is similar to that adopted in \citet{Featherstone2016a} and \citet{Camisassa2022}, but we have extended the definition to now include volumetric averages of the diffusion coefficients, which were not allowed to vary radially in those earlier studies.  

As a consequence of the thermal energy flux and the stellar structure profile being fixed across our internally-heated models, simulations with the same value of $\mathrm{Ra_F}$ but different values of $\alpha$ possess the same volume-averaged values for $\nu$ and $\kappa$.  This results in models with low values of $\alpha$ are the most diffusive near the upper boundary, whereas models with high values of $\alpha$ are least diffusive at the upper boundary.  This effect is illustrated in Figure \ref{fig:viscosity}.

For a smaller subset of our models, we fix the entropy contrast $\Delta S$ across the convection zone.  In these fixed-entropy cases, the non-dimensional buoyancy forcing is controlled by a Rayleigh number based on the entropy contrast, $\mathrm{Ra_S}$, defined as
\begin{equation}
\label{eq:RaS}
    \mathrm{Ra_S} = \frac{\tilde{g} \Delta S H^3}{c_p \widetilde{\nu} \widetilde{\kappa}}. 
\end{equation}
A summary of the input parameters for each model is provided in Tables \ref{tab:internalheatparameters}-\ref{tab:ffparameters}.

\shellfig

\spectrafig

\kefig

\section{Results}

\subsection{Flow Morphology and Distribution}
The central result of this study is summarized in Figure \ref{fig:flow} which illustrates how flow speed and spatial distribution respond to changes in the diffusivity parameter $\alpha$.  We find that, at the same Rayleigh number, the distribution of kinetic energy in depth is correlated with $\alpha$.  Smaller, more negative values of $\alpha$ lead to stronger flows near the lower boundary where viscous effects are weakest.  As $\alpha$ is increased and the viscosity weakens with height, regions of strong flow increasingly shift toward the upper boundary and away from the lower boundary.   

This trend is also apparent when considering the spectral distribution of velocity power, $P_\ell(r)$, defined as
\begin{equation}
\label{eq:power}
P_\ell(r) = \sum_{m=-\ell}^{\ell} \boldsymbol{v}_{\ell}^{m}(r)\cdot\boldsymbol{v}_{\ell}^{*m}(r),
\end{equation}
where $\boldsymbol{v}_{\ell}^m(r)$ is the spherical harmonic transform of the full velocity vector at radius $r$.  Figure \ref{fig:spectra} depicts the velocity power spectra in the lower, middle, and upper convection zone for those models employing internal heating and with $\mathrm{Ra_F} = 10^4$. The depths plotted correspond to $\mathrm{r/R_o}$ = [0.75, 86, 0.98] respectively. Each curve has been time-averaged over at least one viscous diffusion time at the end of each equilibrated model's evolution.   

The trend observed in Figure \ref{fig:flow} is visible in Figure \ref{fig:spectra} as well.  At low values of $\alpha$, small-scale convective power is highest in the lower domain.  As $\alpha$ is increased, small-scale power is enhanced in the upper domain.   A central finding of this study is thus that the parameter $\alpha$ controls the relative distribution of turbulence within the convective shell.

\subsection{Kinetic Energy Scaling}
From a global point of view, we find that the differences resulting from changes in $\alpha$ are considerably more subtle.  This can be seen in Figure \ref{fig:kevraf}, where we have plotted the volume-integrated kinetic energy, KE, defined as
\begin{equation}
\label{eq:}
    \mathrm{KE} = \frac{1}{2}\int_V{\overline{\rho}(r)~|\bm{u}(r, \theta, \phi)|^{2}} \, dV ,
\end{equation}
The integral is carried out over the full shell and subsequently time-averaged over at least one viscous diffusion time following equilibration.   As the value of $\mathrm{Ra}_\mathrm{F}$ is increased, the integrated kinetic energy converges to a value of $\approx 6\times10^{39}$ erg cm$^{-3}$, regardless of the value chosen for $\alpha$.  This is similar to the results reported in \citet{Featherstone2016a}, where models run with different density contrasts across the layer converged to a similar value of KE.   We note that the critical value of $\mathrm{Ra}_\mathrm{F}$ seems to exhibit some weak dependence on $\alpha$, with lower-$\alpha$ models displaying convergence at lower values of the flux Rayleigh number than their high-$\alpha$ counterparts. Note that the high-$\alpha$ series of models has yet to converge for the parameters considered here, likely due to the excess diffusivity at depth. This suggests that the volume-averaged definition of $\mathrm{Ra}_\mathrm{F}$ provided by Equation \ref{eq:RaF} may not perfectly represent the change in free-fall timescale introduced by radially-varying diffusivity.


By examining the Reynolds, number $\widetilde{\mathrm{Re}}$, the similarities between these cases become more apparent. The Reynolds number measures the ratio between inertial and viscous forces, and we define it as

\begin{equation}
        \widetilde{\mathrm{Re}} = \frac{u_\mathrm{rms}H}{\widetilde{\nu}},
\end{equation}
where H is the shell depth, and where the rms average of $u$ is taken over the full convective shell.

\figre

\figreff

The scaling of $\widetilde{\mathrm{Re}}$ versus $\mathrm{Ra_F}$ is shown in Figure \ref{fig:VARe}.  There, we illustrate the results for the internally-heated models shown in Figure \ref{fig:kevraf}, as well as for models that employ fixed-flux and fixed-entropy boundary conditions. For the internally-heated systems, when $ 10^4 \le \mathrm{Ra_F} \lesssim 10^6$, we find that
\begin{equation}
    \mathrm{Re} \propto \mathrm{Ra_F}^{0.337 \pm 0.009}.
\end{equation}
This value of the exponent is close to the so-called ``free-fall'' value of 1/3 which indicates that a leading-order force balance between inertia and buoyancy has been established \citep[e.g.,][]{Featherstone2016a,Orvedahl2018, aurnou2020}.  By contrast, for values of $\mathrm{Ra_F}  \lesssim 10^4 $, the system is near the onset of convection and this scaling does not hold, indicating that diffusion plays a significant role in the global force balance.


These results indicate that models with internal heating achieve the diffusion-free regime even near onset.  Without internal heating, however, this is not the case.  This can be seen by considering the green and violet symbols in Figure \ref{fig:VARe}. Those correspond to models with fixed-entropy and fixed-flux boundary conditions respectively.  Both of curves exhibit a slope greater than 1/3.  

The apparent tendency of the internally-heated models to achieve free-fall scaling at such low Reynolds numbers is somewhat puzzling.   It is possible that some of this behavior arises due to the volume-averaged measurement of $\widetilde{\mathrm{Re}}$ that we have adopted.  In compressible systems such as those considered here, the downflows form a network of plumes that are more rapid, but less space-filling than the upflows.  The result is that the weaker upflows will contribute more heavily to the rms average used when computing $\widetilde{\mathrm{Re}}$.

We investigate this possibility by considering an alternative definition of the Reynolds number that is depth-dependent and based on the the maximum flow amplitude, as opposed to the average, at a given depth.  We denote this Reynolds number by Re$_\mathrm{m}$, which we define as
\begin{equation}
    \mathrm{Re_{m}}(r) = \frac{ {\mathrm{max}\left[\bm{u}(r)\right]}   H}  {\tilde{\nu}} .
\end{equation}

In Figure \ref{fig:reff}, we plot the variation of Re$_\mathrm{m}$ with Ra$_\mathrm{F}$ at three different depths, $\mathrm{r/R_o}$ = [0.75, 0.86, 0.98].  There, we have normalized the value of Re$_\mathrm{m}(r)$ by a Reynolds number based on the free-fall speed at that depth, Re$_\mathrm{ff}$, given by
\begin{equation}
    \mathrm{Re_{ff}}(r) = \frac{u_{ff}(r) H }{\tilde{\nu}}.
\end{equation}
Here, $u_{ff}(r)$ denotes the speed achieved at a given depth by a parcel of fluid freely falling from rest at the upper boundary, such that
\begin{equation}
    u_{ff}(r) = \sqrt{\frac{2}{\overline{\rho}(r)}\int_{R_o}^{r}\frac{\overline{\rho}(r') S(r') g(r')}{c_p}dr'} .
\end{equation}

We find that for nearly all models in this study, the resulting maximum flow speed approaches the free-fall speed within a factor of 10 at all depths.  For those models with lower diffusivity near the outer boundary (models with higher values of $\alpha$), the free-fall speed is approached more closely in the upper convection zone.  For models with low values of $\alpha$, near free-fall speeds are maintained in the deep shell. 

Finally, near the upper boundary, for all models considered, the free-fall speed can be exceeded at sufficiently high $\mathrm{Ra_F}$.  We suspect this is due to the fact that not all downflow plumes originate from rest at the uppper boundary. Pressure or inertial forces associated with converging flows near the surface may provide additional impetus to some downflows that are in turn responsible for determining the value of Re$_\mathrm{m}$.

\subsection{Heat Transport}

We next consider how the integrated energy transport achieved in our models varies in response to changes in $\alpha$.  In our non-rotating, non-magnetic models the radial energy transport is comprised of four distinct fluxes: the kinetic energy flux $F_{KE}$, the viscous flux $F_\nu$, the enthalpy flux $F_e$, and the conductive flux $F_c$. We define these as

\figflux

\begin{equation}
    F_{KE} = \frac{1}{2}\overline{\rho}u_r~|\bm{u}^2| , 
\end{equation}
\begin{equation}
    F_\nu = -(\bm{u} \cdot \bm{\mathcal{D}} ) \cdot \bm{\hat{r}} ,
\end{equation}
\begin{equation}
    F_e = \overline{\rho}c_pu_rT ,
\end{equation}
\begin{equation}
    F_c = \kappa \overline{\rho}\overline{T}\frac{\partial S}{\partial r}.
\end{equation}


We plot these fluxes for a selection of models with the same value of $\mathrm{Ra}_\mathrm{F}$, but with different values of $\alpha$ in Figure \ref{fig:fluxes}.  We have also plotted the effective flux associated with radiative heating, $F_\mathrm{rad}$, defined as
\begin{equation}
F_\mathrm{rad}=\frac{L_\star}{4\pi r^2}-F(r),
\end{equation}
where $L_\star$ and $F(r)$ are defined in Equations \ref{eq:lstar} and \ref{eq:radheat}.  We note that the profile of $F_\mathrm{rad}$ is a constant across all models in this study.  All curves have been integrated over solid angle and, as with KE and $P_\ell$, averaged in time.

These heat flux profiles display depth-dependent variations that are qualitatively similar to the structure observed in the flow strengths and spectral energy distributions presented previously.  For low values of $\alpha$, the convective heat flux and kinetic energy flux are both stronger in the lower half of the domain (left of dashed reference line).  As $\alpha$ is increased, and convective flows become stronger in the upper domain, so too do the convective enthalpy flux and the kinetic energy flux.  We also observe that the thermal boundary layer becomes thicker as $\alpha$ is decreased, as is evident from the conductive flux profiles illustrated in Figure \ref{fig:fluxes}.

\fignu

We can calculate a Nusselt number, and quantify the relative importance of convective and conductive heat transport, by integrating the curves of Figure \ref{fig:fluxes} in depth.  We have chosen to define the Nusselt number in two ways.  The first is an integral definition, with the volume-averaged Nusselt number, $\widetilde{\mathrm{Nu}}$, defined as

\begin{equation}
    \label{eq:Nusselt}
    \widetilde{\mathrm{Nu}} = \frac{\int_V  ( F_{KE} + F_\nu + F_e + F_c) \, dV}{\int  F_c \,  dV}.
\end{equation}
We note that this definition of the Nusselt number differs from that traditionally used in studies based on the Boussinesq approximation.  In particular, it includes the kinetic energy and viscous fluxes and, through the enthalpy flux, contributions due to pressure work.  This is necessary because the thermal and kinetic energy equations do not decouple under the anelastic approximation as they do under the Boussinesq approximation \citep[e.g.,][]{Miesch2005}.

Separately, we can define a Nusselt number based on the upper boundary flux, $\mathrm{Nu}_\mathrm{T}$, as

\begin{equation}
\label{eq:Nuflux}
    \mathrm{Nu_{T}} = \frac{F_{ctop}}{F_{cstate}}.
\end{equation}
Here, $F_{ctop}$ is the conductive flux exiting the upper boundary of the simulation.  $F_{cstate}$ is the total upper-boundary flux for a purely conductive fluid that establishes the same entropy contrast as that resulting from the convective model.  This second formulation has also been used in studies involving spherical anelastic systems \citep{Raynaud2018, Lance2024}.

For either formulation, a Nusselt number of unity indicates that the heat transport is purely conductive in nature, whereas larger values indicate that convection plays an important role in the energy transport.  The scalings with respect to Ra$_\mathrm{F}$ for both types of Nusselt numbers are shown in Fig. \ref{fig:Nu}.  We find that the volume-averaged definition results in higher values of the Nusselt number than the upper-boundary definition.  For either definition, however, the value of Nu-1 is $\mathcal{O}$(1), indicating that thermal conduction plays a significant role in modulating the heat transport.  

The fact that we observe diffusion-free behavior for the kinetic energy and not the heat transport suggests that the bulk of the fluid is effectively inviscid, whereas the heat transfer is ultimately controlled by the thermal boundary layers, which are necessarily diffusive in these models.  This mix of inviscid bulk dynamics with boundary layer limited heat transfer has been observed in the laboratory experiments of \citet{abbate2023, Hawkin2023} and in the numerical simulations of \citet{gastine2016, oliver2023}. 

Diffusion-free behavior has been observed for Nu, however, in the radiatively-driven laboratory experiments of \cite{lepot2018}. There, optical heating deposits thermal energy directly into the fluid bulk, bypassing the throttling effects of any diffusively-controlled thermal boundary layers.  In this case, the diffusivity-free heat transfer scaling naturally manifests.  This has also been explored numerically, with similar results achieved when thermal boundary layers are removed through the use of heating and cooling functions \citep{Lewis2026}.


\section{Conclusions and Perspectives}
In this work, we have examined how variations in the functional form adopted for the kinematic viscosity and thermal diffusivity impact the dynamics and heat transport in a model stellar convection zone.  While the simulations presented here did not account for the effects of rotation and magnetism, they nevertheless hold interesting consequences for stellar convection zone models.

One result of this study is that the integrated kinetic energy of the convection is primarily sensitive to the volumetric average of the diffusion coefficients.   Notably, when the flux Rayleigh number is based on these averaged values, the kinetic energy asymptotically approaches the same constant value as Ra$_\mathrm{F}$ is increased (Figure \ref{fig:kevraf}).  This behavior occurs regardless of the functional form adopted for the diffusion coefficients.  It corresponds to a ``free-fall'' regime of flow in which the dominant force balance is struck between inertia and buoyancy, as evidenced by the scaling of Re with Ra$_\mathrm{F}$ illustrated in Figure \ref{fig:VARe}.  

We have not observed such a diffusion-free regime for the convective heat transport, as illustrated by the scaling of Nusselt number with respect to Ra$_\mathrm{F}$ depicted in Figure \ref{fig:Nu}.  This is true for all values of $\alpha$ considered, as well as for the two sets of models run with fixed-entropy or fixed-flux boundary conditions and no internal heating.  This is not entirely unexpected, however, as the boundary layer in these models is conductive, and so heat transport is invariably modulated by diffusion at the upper boundary.  In fact, previous studies that have observed diffusion-free scaling of the Nusselt number of done so only by effectively removing the thermal boundary layer through the use of a combination of internal heating and cooling \citep{lepot2018,Lewis2026}.

While the integrated kinetic energy of the flow appears to be insensitive to the value of $\alpha$, we do observe significant changes in the morphology and distribution of flow within the convective shell.   As the value of the diffusion exponent $\alpha$ is increased, the region of weakest viscosity and strongest convection shifts from the lower boundary toward the upper boundary of the domain (Figure \ref{fig:flow}).  This effect is accompanied by a change in the distribution of large- and small-scale velocity power with depth such that small-scale velocity power tends to become more concentrated in the lower domain as $\alpha$ is decreased and in the upper domain as $\alpha$ is increased (Figure \ref{fig:spectra}).  The efficiency of convective heat transport similarly shifts in radius with changes in $\alpha$, such that regions of strong and small-scale flow correspond to regions of enhanced convective enthalpy flux (Figure \ref{fig:fluxes}).

This redistribution of velocity power and convective efficiency in response to the form of the diffusivity may have implications for the interpretation of stellar convection models.   For instance, the kinetic helicity of the convection is thought to play a central role in the generation of stellar magnetism through the so-called ``$\alpha$-effect'' \citep[not to be confused with the $\alpha$ employed in this study; e.g.,][]{Mestel1999}.  The distribution of this quantity in depth will likely shift in response to changes in diffusivity in a similar fashion to the small-scale power.  And so, the functional form adopted for the diffusivity may impact the location of strong magnetic field generation.

As a more concrete example, we consider the comparison between helioseismic observations and solar convection-zone models carried out by \citet{Hanasoge2012}.  That study concluded that models evinced considerably higher convective velocity amplitudes than were measurable in the upper 1/3 of the solar convection zone.  The result sparked vigorous debate and questioning, now termed ``the convective conundrum,''  concerning the likely amplitude and structure of deep solar convection \citep[e.g.,][]{Omara2016}.  And yet, the study of \citet{Hanasoge2012} considered a single model with a value of $\alpha=0$.  Had models with alternative formulations for the diffusivity been considered, regions of strong flow may have shifted in radius, possibly leading to a different set of conclusions.  This is not to say that $\alpha=0$ was an incorrect (or correct) choice.  Nor are we questioning the validity of the \citet{Hanasoge2012} result. We are instead pointing out that the prescription for the diffusivities adopted in a particular model can have potentially unexpected implications for any comparisons between models and observations.

\begin{acknowledgements}
B.~Lazard received support through NASA's Future Future Investigators in NASA Earth and Space Science and Technology fellowship award \#80NSSC24K1866. N.~Featherstone received support through NASA grants  \#80NSSC22M0162 (COFFIES DRIVE Center), \#80NSSC24K0125 (HTMS) and \#80NSSC24K0271 (HSR) and NSF grant 2405049 (SHINE). J.~Aurnou was supported by NSF EAR award \#2143939. Computational resources were provided by NASA’s High-End Computing (HEC) program through the Pleiades supercomputer. The Rayleigh code has been developed with support by the National Science Foundation through the Computational Infrastructure for Geodynamics under grants NSF-0949446 and NSF-1550901.  

\end{acknowledgements}

\tableinternal

\tablefixed

\tableflux

\bibliography{paper}{}
\bibliographystyle{aasjournal}
\end{document}

%% file: Tables.tex
\def\tablepoly{
\begin{table}[h]
    \centering
    \renewcommand{\arraystretch}{1.2}
    \begin{tabular}{l l}
        \hline
        \multicolumn{2}{c}{Polytropic Background State Parameters} \\
        \hline
        $r_i$ & $5.000 \times 10^{10}$ cm  \\
        $r_o$ & $6.747 \times 10^{10}$  cm \\
        $M_i$ & $1.989 \times 10^{33} \text{ g}$ \\
        ${\rho}_i$ & $1.805 \times 10^{-1} \text{ g cm}^{-3}$ \\
        $c_p$ & $3.500 \times 10^8 \text{ erg K}^{-1} \text{ g}^{-1}$ \\
        $\gamma$ & 5/3 \\
        $n$ & 3/2 \\
        $N_{\rho}$ & 4 \\
        \hline
    \end{tabular}
        \caption{Summary of system parameters specifying the polytropic background state used here and described in detail in \citet{JONES2011} and \citet{Featherstone2016a}. Indicated are the values of inner radius $r_i$, the outer radius $r_o$, the mass interior to the lower boundary $M_i$, the density at the inner boundary $\rho_i$, the specific heat at constant pressure $c_p$, the adiabatic index $\gamma$, the polytropic index $n$ and the number of density scaleheights across the shell $N_\rho$.}\label{table:poly}
\end{table}
}

\def\tableinternal{
\begin{table*}[p] 
  \centering
  \begin{tabular}{ccccccccccc}
    \toprule
    \multicolumn{4}{c}{Input Parameters} & \multicolumn{7}{c}{Output Parameters} \\
    \cmidrule(r){1-4} \cmidrule(l){5-11}
    $\nu_{12}$ & $\alpha$ & $\mathrm{Ra_{F}}$ & $\ell_{max}$ & KE$_{38}$ & $\mathrm{Ra_{s}}$ & $\mathrm{\widetilde{Nu}}-1$ &$ \mathrm{Nu_{T}}-1$ & $\mathrm{\widetilde{Re}}$ & $\mathrm{Re_{m}}$ & $\mathrm{Re_{ff}}$ \\
    \midrule
    \midrule
    5.923 & 0.5 & $ 10^3$ & 511 & 12.17 & $1.70 \times 10^4$ & 0.535 & 0.183 & 6.90 & 38.72 & 70.30 \\
    4.107 & 0.5 & $3 \times 10^3$ & 511 & 30.63 & $4.53 \times 10^4$ & 2.18 & 0.334 & 14.77 & 57.26 & 90.59 \\
    2.750 & 0.5 & $10^4$ & 511 & 45.08 & $1.23 \times 10^5$ & 4.24 & 0.632 & 25.07 & 95.29 & 129.86 \\
    \midrule
    22.95 & 0 & $10^3$ & 127 & 19.52 & $7.14 \times 10^3$ & 0.995 & 0.076 & 8.51 & 19.24 & 61.55 \\
    15.92 & 0 & $3 \times 10^3$ & 127 & 40.280 & $2.04 \times 10^4$ & 1.70 & 0.129 & 14.81 & 43.42 & 99.73 \\
    10.65 & 0 & $10^4$ & 127 & 54.93 & $5.97 \times 10^4$ & 2.84 & 0.287 & 24.16 & 80.48 & 147.89 \\
    7.387 & 0 & $3 \times 10^4$ & 255 & 57.44 & $1.56 \times 10^5$ & 4.11 & 0.475 & 35.32 & 125.98 & 202.83 \\
    4.945 & 0 & $10^5$ & 255 & 57.82 & $4.36 \times 10^5$ & 6.13 & 0.759 & 54.30 & 197.40 & 293.91 \\
    3.492 & 0 & $3 \times 10^5$ & 511 & 59.17 & $1.09 \times 10^6$ & 8.72 & 1.10 & 74.55 & 332.29 & 416.90 \\
    2.295 & 0 & $10^6$ & 511 & 60.78 & $2.93 \times 10^6$ & 12.75 & 1.61 & 117.34 & 589.97 & 615.50 \\
    \midrule
    67.84 & -0.5 & $10^3$ & 127 & 37.60 & $3.71 \times 10^3$ & 1.11 & 0.231 & 9.16 & 25.55 & 62.90 \\
    47.04 & -0.5 & $3 \times 10^3$ & 127 & 53.02 & $1.05 \times 10^4$ & 1.53 & 0.299 & 14.42 & 40.03 & 102.90 \\
    31.49 & -0.5 & $10^4$ & 127 & 58.83 & $3.20 \times 10^4$ & 2.12 & 0.428 & 21.04 & 67.95 & 156.05 \\
    21.83 & -0.5 & $3 \times 10^4$ & 255 & 59.29 & $8.53 \times 10^4$ & 2.91 & 0.600 & 32.28 & 107.69 & 215.38 \\
    14.62 & -0.5 & $10^5$ & 255 & 59.62 & $2.47 \times 10^5$ & 4.15 & 0.852 & 52.62 & 163.20 & 310.41 \\
     10.13 & -0.5 & $3 \times 10^5$ & 511 & 60.20 & $6.38 \times 10^5$ & 5.63 & 1.15 & 74.73 & 253.91 & 435.54 \\
    6.784 & -0.5 & $10^6$ & 511 & 60.90 & $1.75 \times 10^6$ & 7.72 & 1.61 & 106.17 & 383.27 & 628.05 \\
    \midrule
    149.1 & -1 & $ 10^3$ & 127 & 54.04 & $2.81 \times 10^3$ & 1.29 & 0.709 & 8.78 & 25.35 & 71.36 \\
    103.4 & -1 & $3 \times 10^3$ & 255 & 58.86 & $7.75 \times 10^3$ & 1.63 & 0.864 & 14.09 & 42.24 & 105.29 \\
    69.20 & -1 & $10^4$ & 511 & 61.64 & $1.23 \times 10^5$ & 2.14 & 1.10 & 22.63 & 64.06 & 153.67 \\
    47.98 & -1 & $3 \times 10^4$ & 511 & 62.89 & $6.09 \times 10^4$ & 2.75 & 1.37 & 33.21 & 101.65 & 217.87 \\
    \bottomrule
  \end{tabular}
  \caption{Simulation parameters for models with internal heating.  The dimensional value viscosity at the upper boundary, $\nu_{12}$, is provided in units of 10$^{12}$ cm$^2$ s$^{-1}$. The dimensional value of the volume-integrated kinetic energy, KE$_{38}$, is provided in units of 10$^{38}$ erg.  All models were run with 128 collocation points in radius.}
  \label{tab:internalheatparameters} 
\end{table*}

}

\def\tableflux{
\begin{table*}[t] 
  \centering
  \begin{tabular}{ccccccccccc}
    \toprule
    \multicolumn{4}{c}{Input Parameters} & \multicolumn{7}{c}{Output Parameters} \\
    \cmidrule(r){1-4} \cmidrule(l){5-11}
$\nu_{12}$ & $\alpha$ & $\mathrm{Ra_{F}}$ & $\ell_{max}$ & KE$_{38}$ & $\mathrm{Ra_{s}}$ & $\mathrm{\widetilde{Nu}}-1$ &$ \mathrm{Nu_{T}}-1$ & $\mathrm{\widetilde{Re}}$ & $\mathrm{Re_{m}}$ & $\mathrm{Re_{ff}}$ \\
    \midrule
    \midrule
    10.65 & 0 & $10^4$ & 127 & 2.90 & $1.99 \times 10^3$ & 0.297 & 0.098 & 6.29 & 19.27 & 100.72 \\
    7.387 & 0 & $3 \times 10^4$ & 255 & 6.69 & $6.60 \times 10^3$ & 0.855 & 0.213 & 12.81 & 37.24 & 151.98 \\
    4.945 & 0 & $ 10^5$ & 255 & 11.82 & $2.54 \times 10^4$ & 1.71 & 0.353 & 23.38 & 70.75 & 238.44 \\
    3.429 & 0 & $3 \times 10^5$ & 511 & 15.87 & $8.37 \times 10^4$ & 2.81 & 0.539 & 39.42 & 123.33 & 365.18 \\
    2.295 & 0 & $10^6$ & 511 & 22.34 & $3.37 \times 10^5$ & 4.59 & 0.858 & 69.57 & 264.69 & 596.70 \\
    \bottomrule
  \end{tabular}
  \caption{Simulation parameters for models with fixed entropy boundary conditions.  Viscosity and KE are reported in the same units as Table \ref{tab:internalheatparameters}.  All models were run with 128 collocation points in radius.}
  \label{tab:ffparameters} 
\end{table*}
}

\def\tablefixed{
\begin{table*}[h] 
  \centering
  \begin{tabular}{ccccccccccc}
    \toprule
    \multicolumn{4}{c}{Input Parameters} & \multicolumn{7}{c}{Output Parameters} \\
    \cmidrule(r){1-4} \cmidrule(l){5-11}
$\nu_{12}$ & $\alpha$ & $\mathrm{Ra_{F}}$ & $\ell_{max}$ & KE$_{38}$ & $\mathrm{Ra_{s}}$ & $\mathrm{\widetilde{Nu}}-1$ &$ \mathrm{Nu_{T}}-1$ & $\mathrm{\widetilde{Re}}$ & $\mathrm{Re_{m}}$ & $\mathrm{Re_{ff}}$ \\
    \midrule
    \midrule
    22.95 & 0 & $1.31 \times 10^3$ & 127 & 26.42 & $7.10 \times 10^3$ & 0.866 & 0.080 & 5.63 & 22.07 & 66.43 \\
    15.92 & 0 & $3.94 \times 10^3$ & 127 & 50.23 & $2.02 \times 10^4$ & 1.46 & 0.137 & 13.84 & 51.72 & 97.81 \\
    10.65 & 0 & $1.31 \times 10^4$ & 255 & 72.80 & $6.00 \times 10^4$ & 2.24 & 0.279 & 26.21 & 79.02 & 155.42 \\
    7.387 & 0 & $3.94 \times 10^4$ & 255 & 81.15 & $1.59 \times 10^5$ & 3.02 & 0.445 & 40.41 & 139.41 & 242.83 \\
    4.945 & 0 & $1.31 \times 10^5$ & 511 & 83.59 & $4.49 \times 10^5$ & 4.21 & 0.710 & 64.96 & 213.45 & 392.49 \\
    3.492 & 0 & $3.94 \times 10^5$ & 511 & 88.11 & $1.12 \times 10^6$ & 5.68 & 1.04 & 94.92 & 383.22 & 603.70 \\
    2.295 & 0 & $1.31 \times 10^6$ & 511 & 92.91 & $3.00 \times 10^6$ & 7.82 & 1.55 & 147.81 & 627.34 & 963.88 \\
    \bottomrule
  \end{tabular}
  \caption{Simulation parameters for models without internal heating and fixed-flux boundary conditions.  Viscosity and KE are reported in the same units as Table \ref{tab:internalheatparameters}.  All models were run with 128 collocation points in radius.}
  \label{tab:ffparameters} 
\end{table*}
}

%% file: Figures.tex
\def\viscfig{
\begin{figure}
    \centering
    \includegraphics[width=0.9\linewidth]{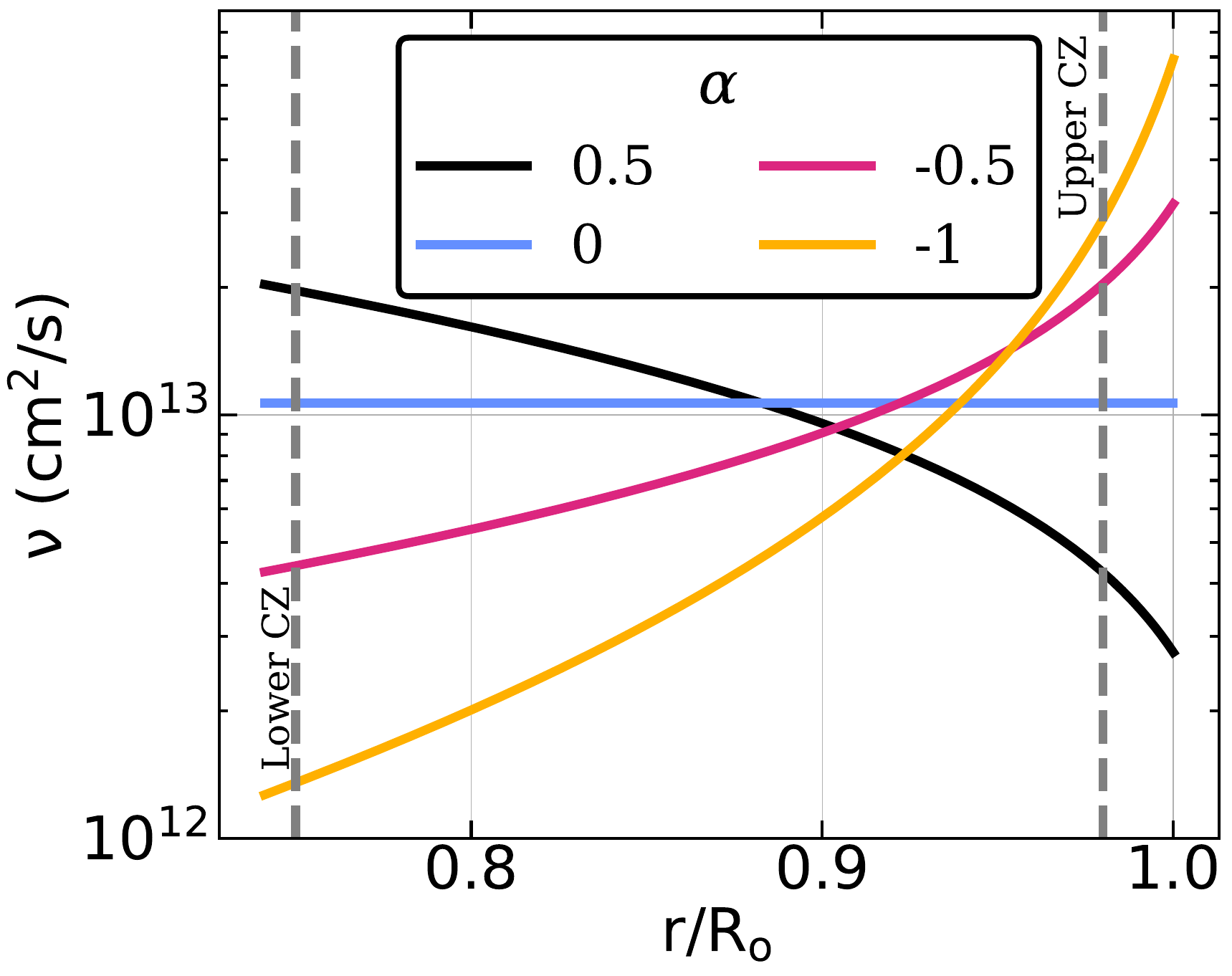}
    \caption{Viscosity versus radius for $\mathrm{Ra_F = 10^4}$. Each curve denotes a different diffusivity parameter $\alpha$ employed in our models. Dashed grey lines indicate at what radial level upper and lower convection zone corresponds that is referenced in this study. For a given $Ra_F$, the volume-averaged viscosity is the same for all models. }
    \label{fig:viscosity}
\end{figure}
}

\def\shellfig{
\begin{figure*}
\centering
\includegraphics[width=0.99\linewidth]{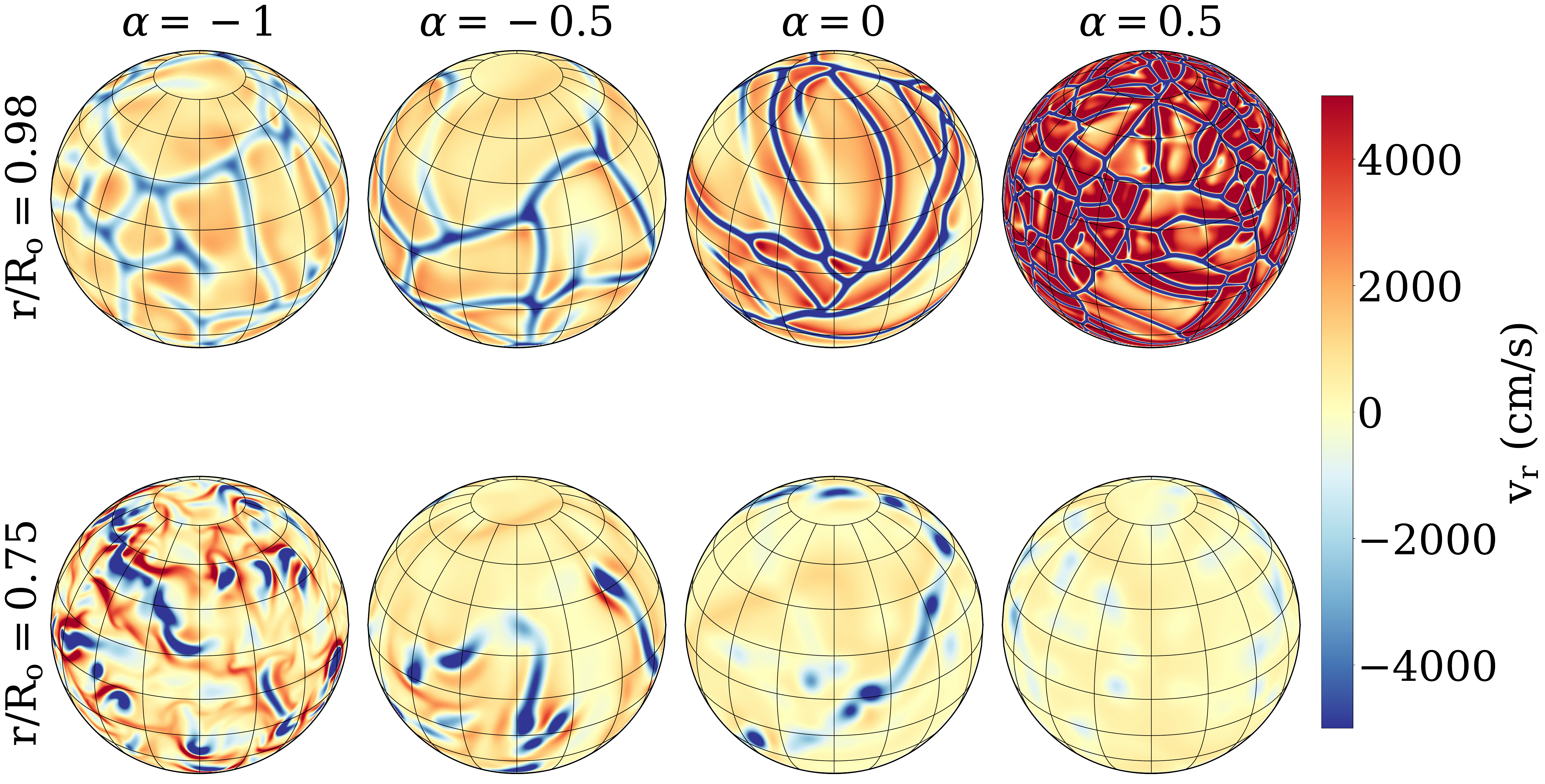}
\caption{Snapshots of radial velocity for $\mathrm{Ra_F = 10^4}$ sampled at a radius of $\mathrm{r/R_o = 0.98}$ (top row) and $\mathrm{r/R_o} = 0.75$ (bottom row).  As $\alpha$ is increased from negative to positive values, the region of strongest convection shifts from the lower convection zone to the upper convection zone.}
\label{fig:flow}
\end{figure*}
}

\def\spectrafig{
\begin{figure*}
    \centering
    \includegraphics[width=0.99\linewidth]{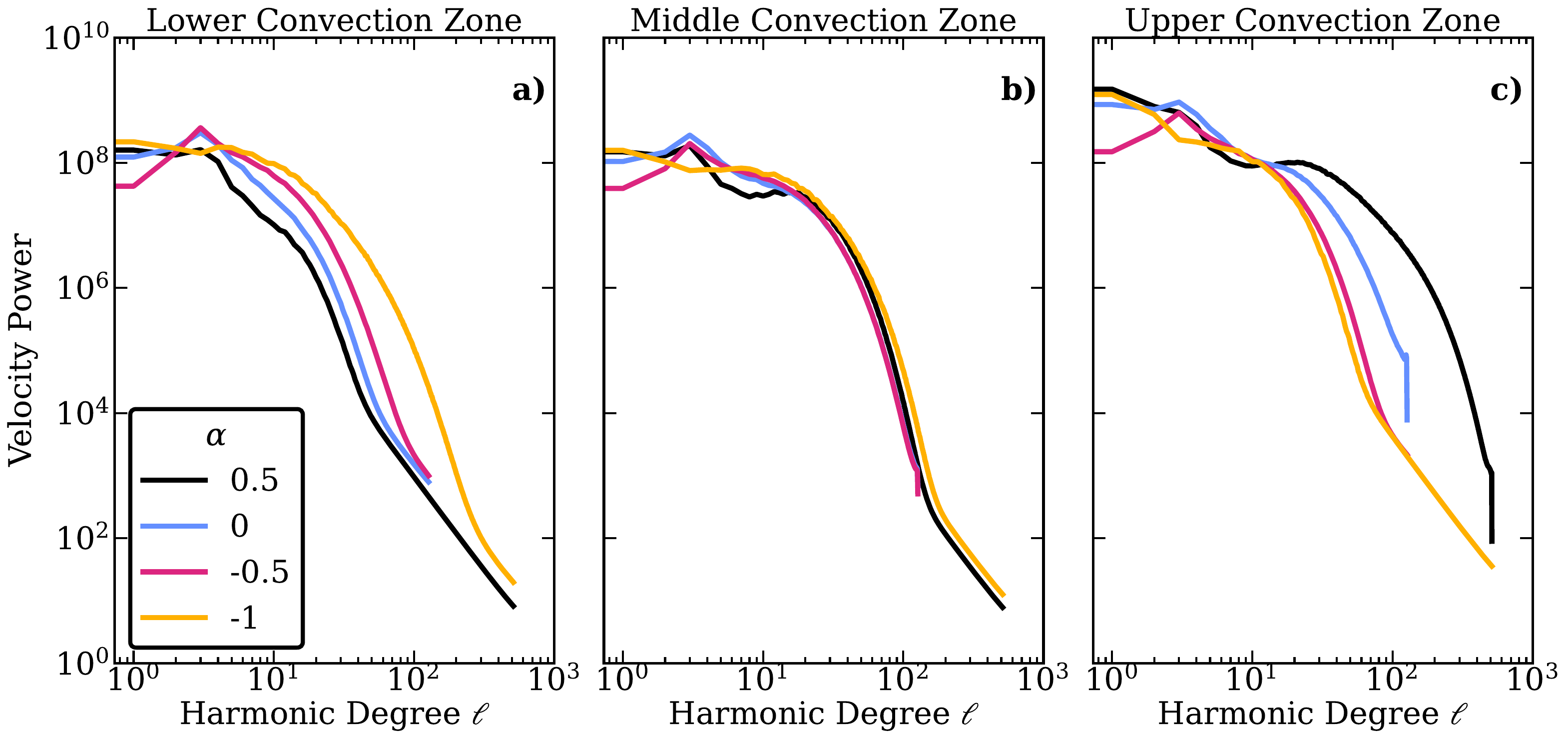}
    \caption{Time-averaged velocity power for $\mathrm{Ra_F =10^4}$ in the lower ($\mathrm{r/R_o = 0.75}$), middle ($\mathrm{r/R_o = 0.86}$), and upper convection zone ($\mathrm{r/R_o = 0.98})$. Each colored curve denotes a different value of $\alpha$. At a given $\mathrm{Ra_F}$, higher values of $\alpha$ result in more small-scale power in the upper convection zone, whereas lower, negative values of $\alpha$ tend to enhance small-scale power in the lower convection zone.}
    \label{fig:spectra}
\end{figure*}
}

\def\kefig{
\begin{figure}
    \centering
    \includegraphics[width=0.99\linewidth]{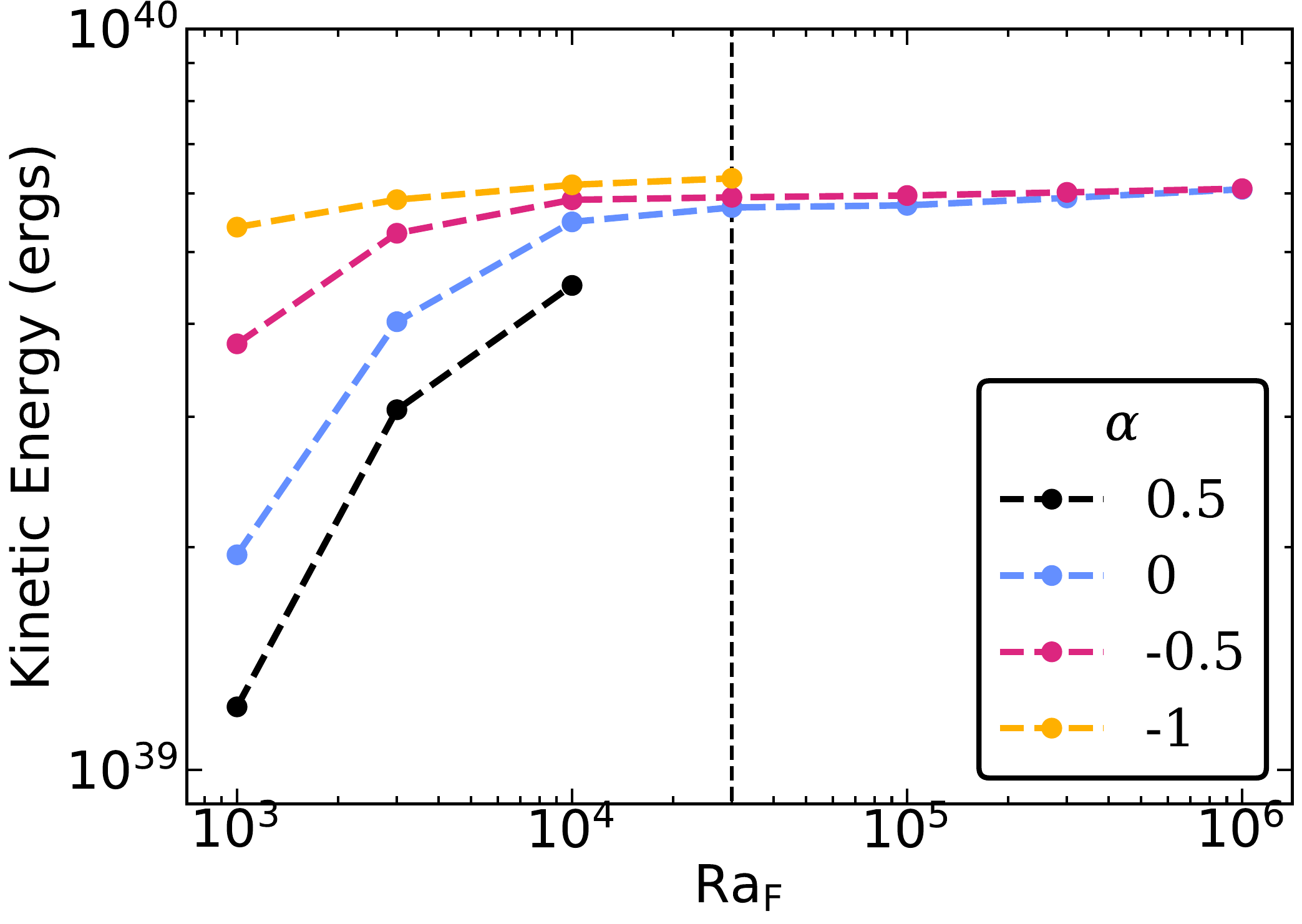}
    \caption{Dependence of dimensional kinetic energy on $\mathrm{Ra_F}$ for internally-heated models. Each colored curve corresponds to a different value of the diffusivity exponent $\alpha$. The dashed line corresponds to the delineation between low-$\mathrm{Ra_F}$ and high-$\mathrm{Ra_F}$ behavior found in \cite{Featherstone2016a}.  At sufficiently high-$\mathrm{Ra_F}$, all models converge toward a similar value of KE, but those with negative values of $\alpha$ asymptote more rapidly than the $\alpha=0$ and $\alpha=0.5$ cases. }
    \label{fig:kevraf}
\end{figure}
}

\def\figre{
\begin{figure}[t]
    \centering
    \includegraphics[width=\linewidth]{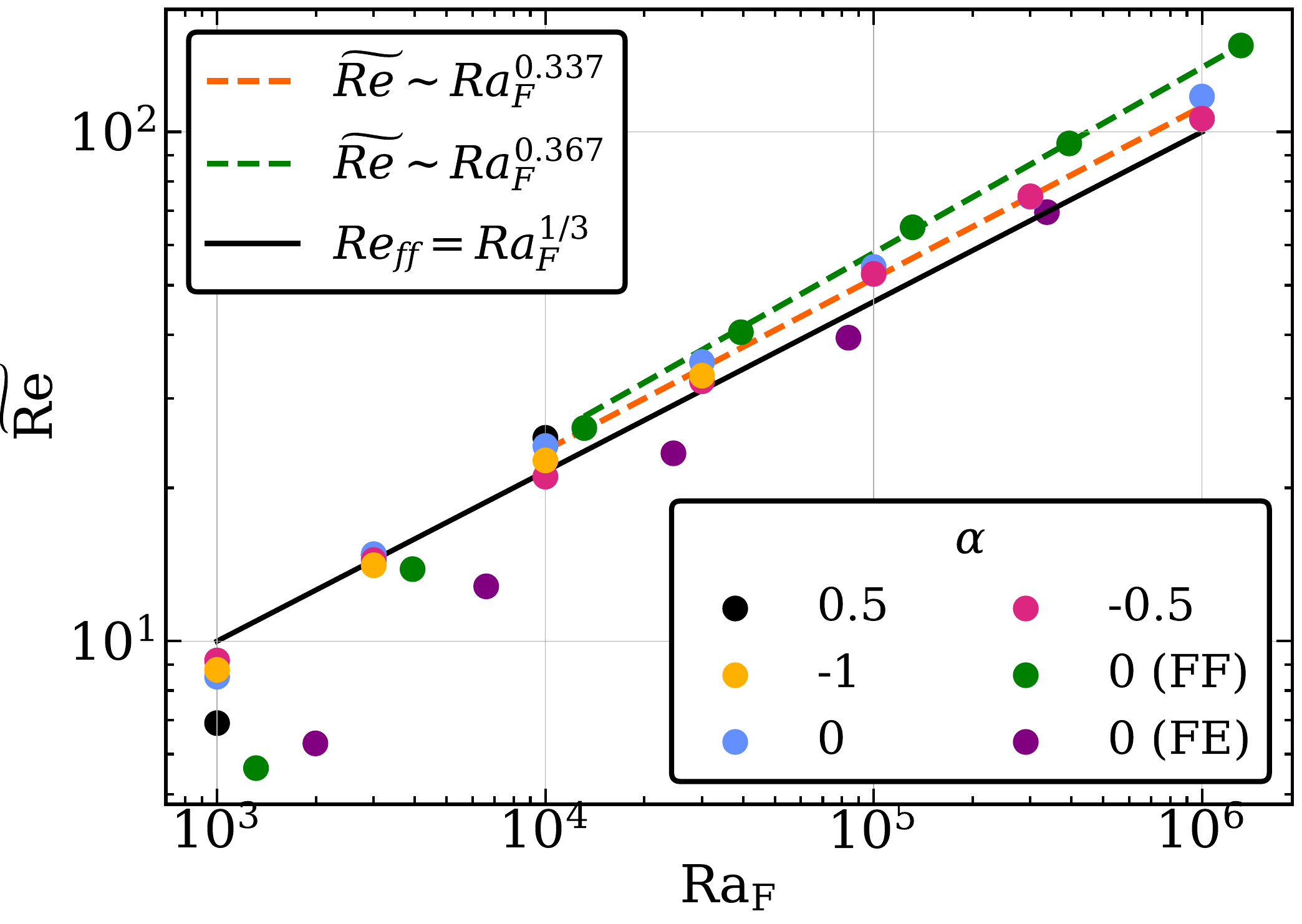}
    \caption{Volume averaged $\mathrm{Re}$ versus $\mathrm{Ra_F}$. The orange line is fit to all internal heating data for $\mathrm{Ra_F \geq 10^4}$. Green dashed line is fit to same range but for fixed flux cases. FF refers to cases without internal heating and fixed flux and FE refers to cases with fixed entropy boundary conditions.}
    \label{fig:VARe}
\end{figure}
}

\def\figreff{
\begin{figure*}
    \centering
    \includegraphics[width=0.99\linewidth]{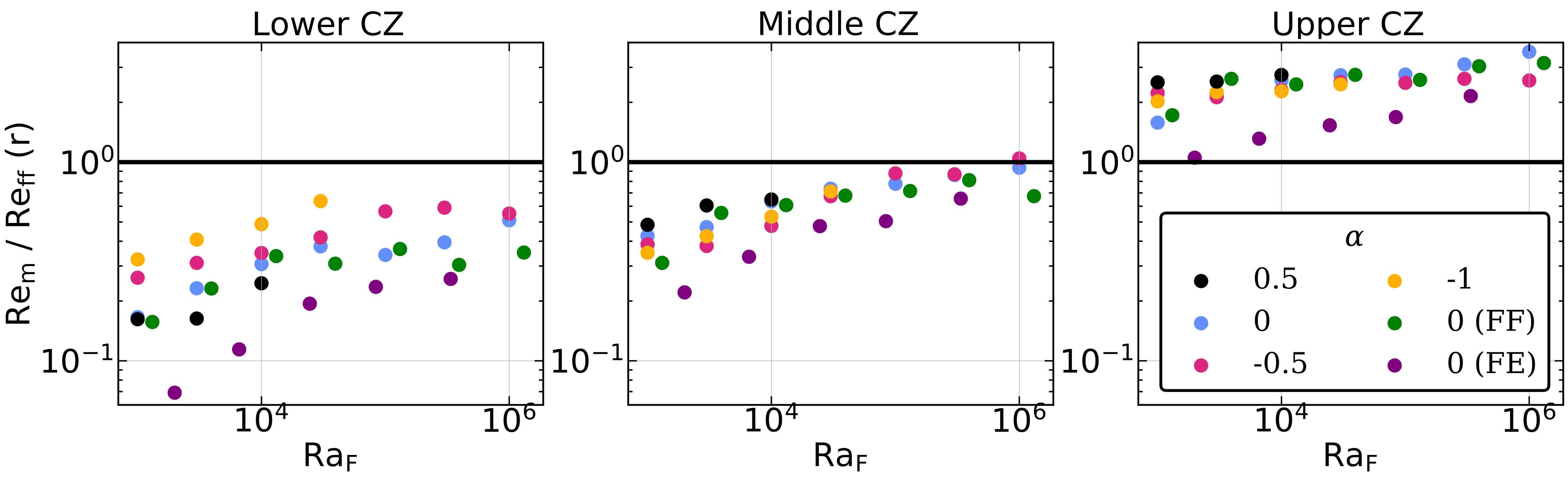}
    \caption{Maximum Reynolds number, $\mathrm{Re}_{m}$, normalized by the free-fall Reynolds number, Re$_\mathrm{ff}$, at three different depths at that corresponding depth (bottom row) versus $\mathrm{Ra_F}$. For models with high values of $\alpha$, the free-fall speed is reached or exceeded in the upper convection zone;  the converse is true for models with low values of $\alpha$.  For the majority of models considered, the maximum resulting flow speed is at least 10$\%$ of the free-fall speed at all depths.}
    \label{fig:reff}
\end{figure*}
}

\def\figflux{
\begin{figure*}
    \centering
    \includegraphics[width=\linewidth]{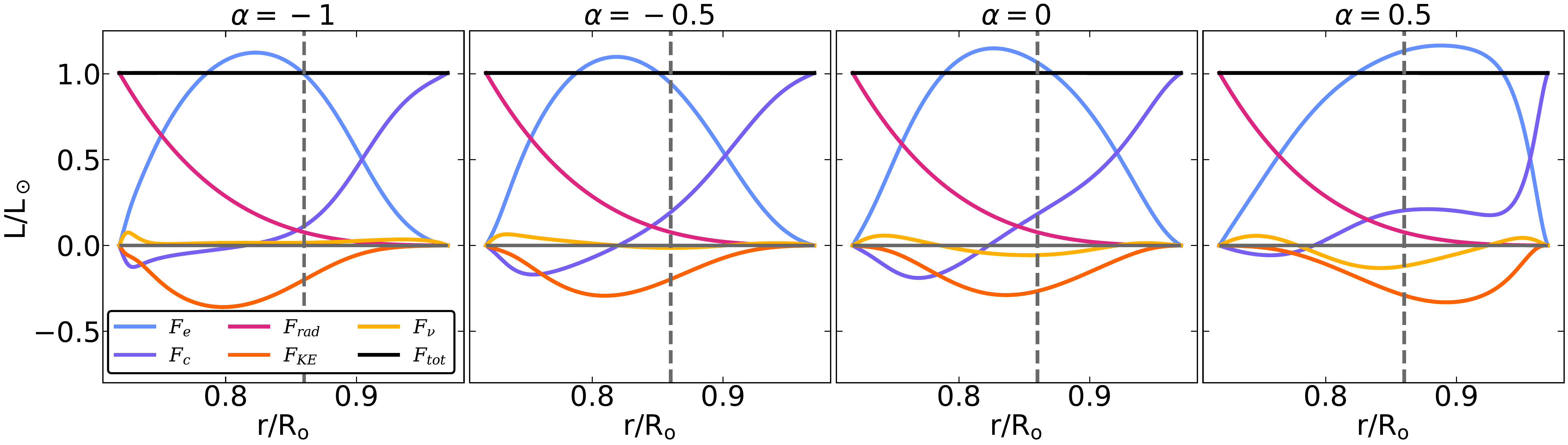}
    \caption{Contributions to the spherically-averaged radial flux balance realized in internally heated models for $\mathrm{Ra_F = 10^4}$.  Each curve has been multiplied by $4\pi r^2$ and normalized by the solar luminosity. The dashed, gray reference line denotes the middle of the convection zone. As $\alpha$ is increased from -1 to 0.5, the region of strong enthalpy and kinetic energy fluxes shifts from the lower convection zone to the upper convection zone.}
    \label{fig:fluxes}
\end{figure*}
}
\def\fignu{
\begin{figure*}
    \centering
    \includegraphics[width=0.9\linewidth]{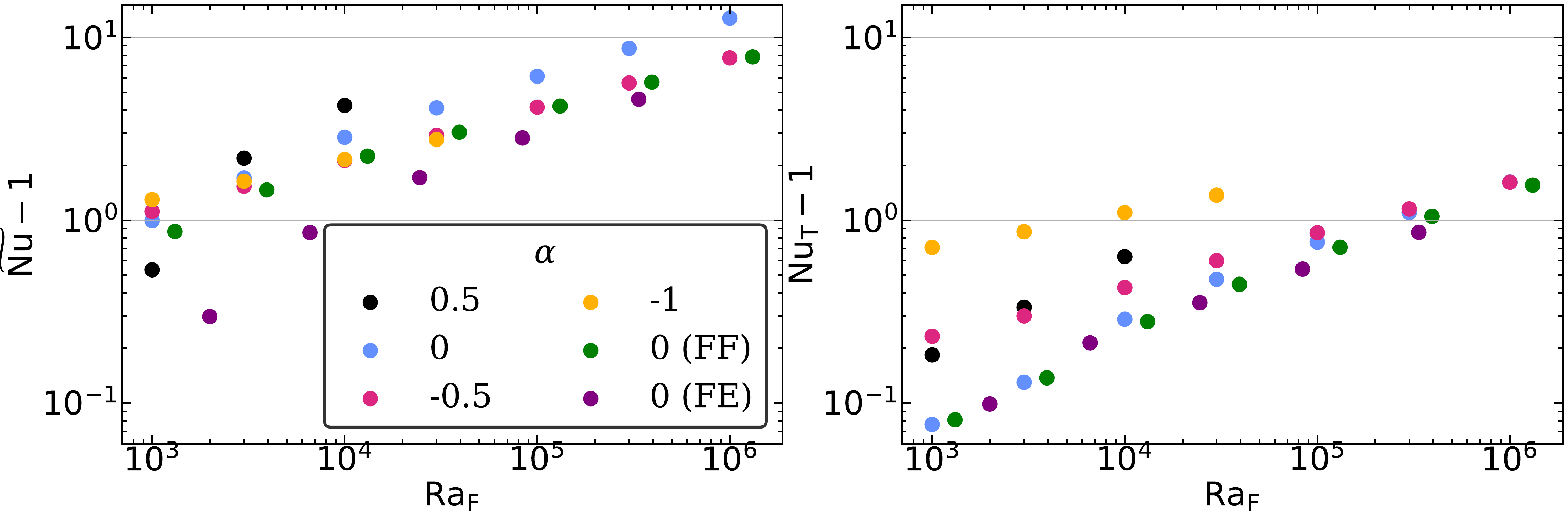}
    \caption{Convective heat flux efficiency for all models considered in this study characterized using the volume-averaged Nusselt number (left panel) and the upper-boundary Nusselt number (right panel). Both formulations yield a value of $\mathcal{O}$(1) for the majority of models considered in this study, suggesting that conduction plays a significant role in the heat transport even at the higher values of Ra$_\mathrm{F}$ employed. }
    \label{fig:Nu}
\end{figure*}
}

%% file: Introduction.tex
\section{Introduction} 
The Sun's magnetic field presumably derives from the motion of plasma flows that pervade its outer, convective envelope. Notably, its axisymmetric east-west and north-south flows, known respectively as differential rotation and meridional circulation, are thought to provide a crucial organizing influence on the large-scale structure and long-time evolution of solar magnetism \citep[e.g., ][]{Charbonneau2020}.  Underpinning these global-scale flows are the Sun's non-axisymmetric, convective motions that convect heat from the radiative interior to the photosphere.  While a variety of observations provide dynamo models with strong constraints on the structure and amplitude of differential rotation and meridional circulation throughout the convection zone  \citep[e.g.,][]{Howe2009,Hanasoge2022}, this is not true for the convection itself.  In this paper, our aim is to explore how the structure and amplitude of convective flows in a model solar convection zone responds to the treatment of subgrid-scale diffusion in those models.

Solar convection is readily observed in the photosphere.  Motions there are dominated by the radiatively-driven solar granulation (1 Mm spatial scale; 1 km s$^{-1}$ flow speed) \citep[e.g., ][]{Herschel1801,Nordlund2009,Hathaway2015} and supergranulation (30 Mm spatial scale; $\approx 400 $ m s$^{-1}$ flow speed) \citep[e.g., ][]{Hart1954,Hart1956,Leighton1962,Rast2003,Goldbaum2009,Langfellner2018,Rincon2018}. Granulation and supergranulation are, however, thought to be too rapid and disorganized to sustain the Sun's differential rotation.  

Instead, theory and direct numerical simulations of rotating convection indicate that a differential rotation consistent with that of the Sun, characterized by a rapidly-rotating equator and slowly-rotating poles, is expected to result from convection that is considerably slower.  Specifically, the overturning time of the convection should be longer than the Sun's rotation period \citep[e.g.,][]{Gastine2014,Featherstone2015}.  Numerical simulations of solar convection indicate that convective flows with speeds of order 10--100 m s$^{-1}$ and horizontal scales comparable to the depth of the convection zone are capable of producing a differential rotation commensurate with that observed in the Sun \citep{Brun2002,Miesch2012,Featherstone2015,Hotta2022}.

In the photosphere, motions on scales larger than granulation and supergranulation are indeed observed \citep[e.g.,][]{Hindman2004, Hathaway2013, Hathaway2021}.  However, these large-scale motions are considerably weaker than estimates based on theory and simulation would suggest and are now understood to be the surface manifestation of global-scale inertial waves \citep[e.g.,][]{Gizon2021,Triana2022,Waidele2023,Hanson2024}.  At depth, the structure and amplitude of convective flows is much less clear.  Inferences resulting from different formulations of helioseismic analysis yield results that can disagree by orders of magnitude on the largest spatial scales \citep{Hanasoge2012,Greer2015,Proxauf2021,Birch2024}.

Numerical simulations of solar convection offer the possibility of resolving these inconsistencies, and yet those efforts yield similarly conflicting results.  For instance, 3-D simulations that successfully reproduce the solar differential rotation tend to possess deep flows with speeds of $\mathcal{O}$(100) m s$^{-1}$ \citep{Brun2002,Miesch2006,Hotta2015}.  Magnetism that exhibits coherence on large-scales and which evinces cyclic behavior reminiscent of the Sun appears to be correlated, however, with simulated flow speeds which are an order of magnitude lower \citep[e.g.,][]{Ghizaru2010, Racine2011,Brown2011,Kapyla2012}.  

Moreover, as solar convection models are pushed toward more turbulent regimes by lowering their viscosity and thermal diffusivity, they tend to develop a differential rotation profile characterized by a slowly-rotating equator and rapidly-rotating poles \citep{Gilman1977, Aurnou2007, Guerrero2013, Gastine2014, Kapyla2014, Featherstone2015, Camisassa2022, Hindman2020, soderlund2025}.  The solution to this particular problem may lie in the effects of small-scale magnetism which, in the models of \citet{Hotta2022} and \citet{Hotta2025}, generates Lorentz forces and torques that are capable of reversing this trend.  Those models do not, however, generate significant large-scale magnetic fields that cycle similarly to that of the Sun. Meanwhile, large-scale magnetism now appears to be a crucial ingredient for sustaining a tachocline \citep{Matilsky2022, Matilsky2024, Matilsky2025}.

At the heart of these discrepancies among numerical models lies an array of different  approaches taken when accounting for the effects of subgrid-scale diffusion (i.e., diffusive processes occurring on spatial scales too small to model directly).   Many models have employed a large-eddy formulation, invoking an explicit diffusion operator, albeit with diffusion coefficients that are substantially enhanced with respect to their solar values \citep[e.g., ][]{Brun2002,Kapyla2011,Featherstone2015}.  Others, meanwhile, allow dissipation to be handled implicitly by the numerical scheme, with the grid resolution effectively setting the amplitude of the dissipation coefficients \citep[e.g.,][]{Ghizaru2010,Guerrero2013,Hotta2025}.

In this study, we examine how the formulation of the diffusion scheme can impact the properties of a simulated stellar convection zone.  In particular, we consider a large-eddy formulation of the viscous and thermal diffusivities, where the diffusion operators are explicitly included in the momentum and thermal energy equations.    Across a suite of otherwise identical models, we vary the amplitude of the diffusion coefficients as well as their variation with depth across the convective layer.  As we do so, we explore the response of the convection's kinetic energy amplitude, its distribution in depth and across wavenumber, and implications for the efficiency of heat transport.